\documentclass[12pt]{article}
\usepackage[pctex32]{graphics}
\begin{document}
\vspace{2cm}
\begin{center}
~
\\
~
{\bf  \Large AdS/CFT Approach to Melvin Field Deformed Wilson Loop}
\vspace{1cm}

                      Wung-Hong Huang\\
                       Department of Physics\\
                       National Cheng Kung University\\
                       Tainan,Taiwan\\

\end{center}
\vspace{2cm}
We first apply the transformation of mixing azimuthal and internal coordinate to the 11D M-theory with a stack N M2-branes to find the spacetime of a stack of  N D2-branes with Melvin one-form in 10D IIA string theory, after the Kaluza-Klein reduction.  Next, we apply the Melvin twist to the spacetime and perform the T duality to obtain the background of a stack of  N D3-branes.  In the near-horizon limit the background becomes the Melvin field deformed $AdS_5 \times S^5$ with NS-NS B field.    In the AdS/CFT correspondence it describes a non-commutative gauge theory with non-constant non-commutativity in the Melvin field deformed spacetime.   We study the Wilson loop therein by investigating the classical Nambu-Goto action of the corresponding string configuration and find that, contrast to that in the undeformed spacetime, the string could be localized near the boundary.  Our result shows that while the geometry could only modify  the Coulomb type potential in IR there presents a minimum distance between the quarks.  We argue that the mechanism behind producing a minimum distance is coming from geometry of the Melvin field deformed background and is not coming from the space non-commutativity.

\vspace{3cm}
\begin{flushleft}
*E-mail:  whhwung@mail.ncku.edu.tw\\
\end{flushleft}


\newpage
\section{Introduction}
The expectation value of Wilson loop is  one of the most important observations in the gauge theory.   In the AdS/CFT duality [1-3]  it becomes tractable to understand this highly nontrivial quantum field theory effect through
a classical description of the string configuration in the AdS background.  Using this AdS/CFT duality Maldacena [4]  derived for the first time the expectation value of the rectangular Wilson loop operator from the Nambu-Goto action associated with the $AdS_5 \times S^5$ supergravity and found that the interquark potential exhibits the Coulomb type behavior expected from conformal invariance of the gauge theory.

Maldacena's computational technique has already been extended to the finite temperature case by replacing the AdS metric by a Schwarzschild-AdS metric [5].  In order to make contact with Nature many investigations had gone beyond the initial conjectured duality and generalized the method to investigate the theories breaking conformality and (partially)  supersymmetry [6-9].  For example, the Klebanov--Witten solution [10], the Klebanov--Tseytlin solution [11], the Klebanov--Strassler solution [12], and Maldacena--N\'u\~nez (MN) solution [13] which dual to the ${\cal N}=1$ gauge theory.

  The technique has also been used to investigate the Wilson loop in non-commutative gauge theories. First,  the dual supergravity description of non-commutative gauge theory with a constant non-commutativity was constructed by Hashimoto and Itzhaki [14].   The corresponding string background is the curved spacetime where the constant B field is turned on along the brane worldvolume.  The associated Wilson loop studied by Maldacena and Russo [15] showed that the string cannot be located near the boundary and we need to take a non-static configuration in which the quark-antiquark acquire a velocity on the non-commutative space.  The result shows that the interquark potential exhibits the Coulomb type behavior as that in commutative space. 

 In [16]  we  use AdS/CFT correspondence to investigate the Wilson loop on the non-commutative gauge theory with a non-constant non-commutativity.  The dual supergravity is the Melvin-Twist deformed $AdS_5\times S^5$ background which was first constructed by Hashimoto and Thomas [17]. The corresponding string is on the curved background where the non-constant B field is turned on along the brane worldvolume.  After analyzing the Nambu-Goto action of the classical string configuration we had shown that [16], while the non-commutativity could modify  the Coulomb type potential in IR it may produce a strong repulsive force between the quark and anti-quark if they are close enough.  In particular, we show that there presents a minimum distance between the quarks, which is proportional to the value of the non-commutativity.  In this paper we will extend the previous paper to study the non-commutative gauge theory in a Melvin field deformed spacetime in the dual string description.

In section II we first apply the transformation of mixing azimuthal and internal coordinate [18] to the 11D M-theory with a stack N M2-branes [19] to find  the spacetime of a stack of  N D2-branes with Melvin one-form in 10D IIA string theory, after the Kaluza-Klein reduction.  Next, we apply the Melvin twist to the spacetime and perform the T duality [20] to obtain the background of a stack of  N D3-branes.  In the near-horizon limit the background becomes the Melvin field deformed $AdS_5 \times S^5$ with NS-NS field.    In the AdS/CFT correspondence it describes the non-commutative gauge theory  in the Melvin field deformed spacetime.

In section III we investigate the Wilson loop therein by dual string description.  We find that, contrast to that in the undeformed space, the string could be localized near the boundary.  Our result shows that there presents a minimum distance between the quarks.   We argue that the mechanism of producing a minimum distance is coming from geometry of the Melvin field deformed background, contrast that in [16] which is coming from the space non-commutativity.   In section IV we discuss the particle trajectory on the Melvin field deformed $AdS_5\times S^5$ background and see that the property of the Melvin field effect on the Wilson loop is, more or less, similar to that on the particle trajectory.  The last section is devoted to a short discussion. 


\section{Supergravity Dual of  Non-commutative Gauge theory on Melvin Field Deformed Spacetime}
We first  apply the transformation of mixing azimuthal and internal coordinate [18] to the 11D M-theory with a stack $N$ M2-branes [19] and then apply the Kaluza-Klein reduction to find the spacetime of a stack of  N D2-branes with one-form in 10D IIA string theory.

The full $N$ M2-branes solution is given by
$$ds^2_{11}=H^{-2\over3}\left(-dt^2+dr^2+ r^2d\phi^2\right)+H^{1\over3}\left(dz^2 + dU^2+ U^2 d\Omega_5^2+dx_{11}^2\right),\eqno{(2.1)}$$
$$A^{(3)} = H^{-1} dt\wedge dx_1\wedge dx_2.\hspace{7.7cm}\eqno{(2.2)}$$
$H$ is the harmonic function defined by 
$$ H = 1+ {R\over r^{D-p-3}}, ~~~~~~~r^2\equiv z^2+ \rho^2+x_{11}^2 , ~~~~R \equiv {16\pi G_D\,T_p \,N\over D-p-3 },\eqno{(2.3)}$$
in which  $G_D$ is the D-dimensional Newton's constant and $T_p$ the p-brane tension. In the case of (2.1), $D=11$ and $p=2$.

We first transform the angle $\phi$ by mixing it with the compactified coordinate $x_{11}$ in the following substituting
$$\phi \rightarrow \phi + C x_{11}.\eqno{(2.4)}$$ 
Using the above substitution the line element (2.1)  becomes
$$ds_{11}^2 = H^{-2\over3} \left(- dt^2+ dr^2 \right) + H^{1\over3}\left(dz^2+dU^2 + U^2 d\Omega^2_5  \right) + \left(r^2H^{-2\over3} -{C^2r^4H^{-4\over3}\over H^{1\over3}+ C^2 r^2 H^{-2\over3}}\right)d\phi^2$$
$$+ \left(H^{1\over3} + C^2 H^{-2\over3}\right)\left[dx_{11} + {Cr^2H^{-2\over3} d\phi\over H^{1\over3}+ C^2 r^2 H^{-2\over3}}\right]^2. \eqno{(2.5)}$$
Using the relation between the 11D M-theory metric and string frame metric, dilaton field and Melvin 1-form potential
$$ds_{11}^2= e^{-2\phi/3}ds_{10}^2+  e^{4\phi/3} (dx_{11}+2 A_\mu dx^\mu )^2 , \eqno{(2.6)} $$
the 10D IIA background is described by
$$ds_{10}^2 =\sqrt{1+ C^2 r^2 H^{-1}} \left[H^{-1\over2}\left(- dt^2+ dr^2+{r^2 d\phi^2\over 1+ C^2 r^2 H^{-1}} \right) +H^{1\over2} \left(dz^2+dU^2 +U^2 d\Omega_5^2\right)\right].\eqno{(2.7)}$$ 
$$A_{\phi} ={Cr^2\over 2\left(H+C^2 r^2\right)},\hspace{11cm}\eqno{(2.8)}$$
in which $A_{\phi}$ is the RR one-form potential.   The dilaton field and  other RR potential arisen from RR there-form $A^{(3)}$  will be neglected hereafter as they are irrelevant to our calculation of the Wilson loop in section III.  In the case of  $C=0$  the above spacetime becomes the well-known geometry of a stack of N D2-branes.  Thus, the background describes the spacetime of a stack of  N D2-branes with Melvin field flux.

We next transform the angle $\phi$ by mixing it with the coordinate $z$ in the following substituting
$$\phi \rightarrow \phi + C z.\eqno{(2.9)}$$ 
Using the above substitution the line element (2.7)  becomes
$$ds_{10}^2 =\sqrt{1+ C^2 r^2 H^{-1}} \left[H^{-1\over2}\left(- dt^2+ dr^2+{r^2 d\phi^2\over 1+ C^2 r^2 H^{-1}} \right) + {2B r^2 H^{-1\over2}d\phi dz \over 1+ C^2 r^2 H^{-1}}\right.$$
$$\left.+ \left({B^2 r^2 H^{-1\over2}\over 1+ C^2 r^2 H^{-1}}+ H^{1\over2}\right)dz^2+H^{1\over2} \left(dU^2 + U^2 d\Omega_5^2\right)\right].\eqno{(2.10)}$$ 
 Finally, we perform the T-duality transformation [20] on the coordinate $z$.  The result supergravity background becomes
$$ds_{10}^2 =\sqrt{1+ C^2 r^2 H^{-1}} \left[H^{-1\over2}\left(- dt^2+ dr^2+{r^2 d\phi^2+dz^2\over 1+ C^2 r^2 H^{-1}+ B^2 r^2 H^{-1}} \right) + H^{1\over2}(dU^2 + U^2 d\Omega_5^2)\right].\eqno{(2.11)}$$ 
$$A_{\phi z} ={Cr^2\over 2\left(H+C^2 r^2\right)},~~~~~~~~B_{\phi z} = {B r^2 H^{-1}\over 1+ C^2 r^2 H^{-1}+ B^2 r^2 H^{-1}},\eqno{(2.12)}$$ 
in which $A_{\phi z}$ is the RR two-form potential and $B_{\phi z}$ the NS-NS B field.
\\

Using above two equations we have two results:

(1) In the near-horizon limit Eqs. (2.11) and (2.12) become
$$ds_{10}^2 =\sqrt{1+ C^2 r^2 U^4} \left[U^2\left(- dt^2+ dr^2+{r^2 d\phi^2+dz^2\over 1+ C^2 r^2 U^4+ B^2 r^2 U^4} \right) + U^{-2}(dU^2 + U^2 d\Omega_5^2)\right].\eqno{(2.13)}$$ 
$$A_{\phi z} ={Cr^2 U^4\over 2\left(1+C^2 r^2 U^4\right)},~~~~~~~~B_{\phi z} = {B r^2 U^4\over 1+ C^2 r^2 U^4+ B^2 r^2 U^4}.\eqno(2.14)$$
This is the Melvin field deformed $AdS_5 \times S^5$ with NS-NS B field.  The Melvin field here means the RR two-form potential $A_{\phi z}$ which arises from the Melvin twist of (2.4).  The NS-NS B field is $B_{\phi z}$ which arises from the Melvin twist of (2.9).

(2) Taking $H =1$ then Eqs. (2.11) and (2.12) become
$$ds_{10}^2 =\sqrt{1+ C^2 r^2} \left[\left(- dt^2+ dr^2+{r^2 d\phi^2+dz^2\over 1+ C^2 r^2 + B^2 r^2} \right) + (dU^2 + U^2 d\Omega_5^2)\right].\eqno{(2.15a)}$$ 
$$B_{\phi z} = {B r^2\over 1+ C^2 r^2+ B^2 r^2}.\eqno(2.15b)$$
Using the above IIB background we can find a gauge theory by applying the mapping of Seiberg and Witten [21]
$$\left(G+ \theta\right)^{\mu\nu}= \left[(g+B)_{\mu\nu}\right]^{-1}.\eqno(2.16)$$
The dual 4D gauge theory is thus on the Melvin field deformed spacetime with the line element
$$G_{\mu\nu}dx^\mu dx\nu = \sqrt{1+C^2 r^2}\left[-dt^2+ dr^2+ {r^2 d\phi^2+ dz^2\over 1+ C^2 r^2}\right],\eqno(2.17a)$$
with space non-commutativity
$$\theta^{\phi z} = B.\eqno(2.17b)$$
When $C=0$ above result is just that described by  Hashimoto and Thomas [17].   Note that replacing the coordinate $(t, r,\phi,z)$ by  Cartesian coordinates $(t,x,y,z)$ the  noncommutativity become $\theta^{xz}= - yB$ and $\theta^{yz}= x B $ [17].  Our background thus describes the non-commutative gauge theory on the Melvin field deformed spacetime with a non-constant non-commutativity.

\section{Wilson Loop of  Non-commutative Gauge theory on Melvin Field Deformed Spacetime}
\subsection{Formulation}
Following the Maldacena's computational technique the Wilson loop of a quark anti-quark pair is calculated from a dual string.  The string lies along a geodesic with endpoints on the $AdS_5$ boundary representing the quark and anti-quark positions.  The ansatz for the background string we will consider is 
$$ t=\tau,~~~z=\sigma,~~~U=U(\sigma),\eqno{(3.1)}$$
and rest of the string position is constant in $\sigma$ and $\tau$.  We choose $r=r_0=1$ for a convenience.  The Nambu-Goto action becomes
$$S= {T\over 2\pi}\int d\sigma \sqrt{(1+C^2U^4) (\partial_\sigma U)^2+{U^4(1+C^2 U^4)\over 1+ C^2 U^4+B^2 U^4}},\eqno{(3.2)}$$
in which $T$ denotes the time interval we are considering and we have set $\alpha'=1$.  As the associated Lagrangian $({\cal L})$ does not explicitly depend on $\sigma$ the relation $(\partial_\sigma U){\partial{\cal L}\over \partial(\partial_\sigma U)} - {\cal L}$ will be proportional to an integration constant $U_0$.  This implies the following relation
$${{U^4(1+C^2 U^4)\over 1+ C^2 U^4+B^2 U^4} \over \sqrt{(1+C^2U^4) (\partial_\sigma U)^2+{U^4(1+C^2 U^4)\over 1+ C^2 U^4+B^2 U^4}}}= constant.\eqno{(3.3)}$$
To describe a quark pair with a finite distance the dual string configuration shall satisfy the following boundary condition
$$\partial_\sigma U \rightarrow \infty, ~~~~~as~~~~~ U \rightarrow \infty.\eqno{(3.4)}$$ 

In the case of $C=0$ above condition implies that the $\it constant$  in (3.3) is zero and we have only null solution of $U=0$.   This is what Maldacena and Russo [15] had showed that the string cannot be located near the boundary and we need to take a non-static configuration in which the quark-antiquark acquires a velocity on the non-commutative space.  The case of $C=0$ had been analyzed by us in a previous paper [16], in which we considered the case of  $\phi = v t$.  The dual string thus is  moving with a constant angular velocity $v$.

In the case of $C \ne 0$ the $\it constant$ in (3.3) could be a finite value which may be calculated from another condition
$$\partial_\sigma U =0, ~~~~~as~~~~~ U =U_0,\eqno{(3.5)}$$ 
in which $U_0$ is the minimum of $U$ of the dual string configuration.   In this case (3.3) implies
$${dU\over d\sigma}  = \pm{U^2\over U_0^2} {1\over \sqrt{1+(C^2+B^2) U^4}}{\sqrt{U^4~{1+C^2U^4\over 1+(C^2+B^2) U^4}{1+(C^2+B^2) U_0^4\over 1+C^2 U_0^4}-U_0^4}}.\eqno{(3.6)}$$
As we put the quark at place $z=\sigma =-L/2$ and the anti-quark at $z=\sigma = L/2$ we can use the above relation to find the following relation between the integration constant $U_0$ and $L$.   
$${L/2} = \int_0^{L/2} d\sigma = \int_{U_0}^\infty dU (\partial_\sigma U)^{-1}
={1\over U_0}\int_0^1 {dx \over x^{3\over4}}{\sqrt{x+ (C^2+B^2)U_0^4}\over\sqrt{{x+ C^2U_0^4\over x+ (C^2+B^2)U_0^4}{1+ (C^2+B^2)U_0^4\over 1+ C^2U_0^4} -x}}.\eqno{(3.7)}$$
In the same way, using (3.6) the interquark potential evaluated from the action (3.2) becomes
$$H = {U_0\over 4\pi}\left(\int_1^\infty dy\left[{\sqrt{y^4{1+ C^2U_0^4y^4\over 1+ (C^2+B^2)U_0^4y^4}{1+ (C^2+B^2)U_0^4\over 1+ C^2U_0^4}}\over\sqrt{y^4{1+ C^2U_0^4y^4\over 1+ (C^2+B^2)U_0^4y^4}{1+ (C^2+B^2)U_0^4\over 1+ C^2U_0^4}-1}}y^\epsilon - y^\epsilon \right] -1\right) $$
$$= {U_0\over 4\pi}\int_0^1 dx  x^{-({7\over4}+{\epsilon\over4})} \sqrt{x+C^2U_0^4}{\sqrt{{x+ C^2U_0^4\over x+ (C^2+B^2)U_0^4}{1+ (C^2+B^2)U_0^4\over 1+ C^2U_0^4}}\over\sqrt{{x+ C^2U_0^4\over x+ (C^2+B^2)U_0^4}{1+ (C^2+B^2)U_0^4\over 1+ C^2U_0^4} -x}},\eqno{(3.8)}$$
in which we have followed the prescription of Maldacena [4] by multiplying the integration a factor  $y^\epsilon$ and subtraction the regularized mass of W-boson to find the finite result.   We can use the Eqs.(3.7) and (3.8) to analyze  the  string configuration on the Melvin field deformed background with NS-NS $B$ field.    The result could be used to find the interquark potential for the system on the deformed background with a non-commutativity.

\subsection{Analyses: Interquark Distance}
For a clear illustration we show in figure 1 the function $L(U_0)$ which is found by performing the numerical evaluation of (3.7) for the cases of $B=0.1$ with  $C=0.01$ and $C=0.1$ respectively.   Figure 2 shows the function $L(U_0)$ for the cases of  $C=0.01$ with  $B=0.1$ and $B=0.2$ respectively. The dashed line therein represents that with $C=B=0$.  
\\
\\
\scalebox{1}{\hspace{5cm}\includegraphics{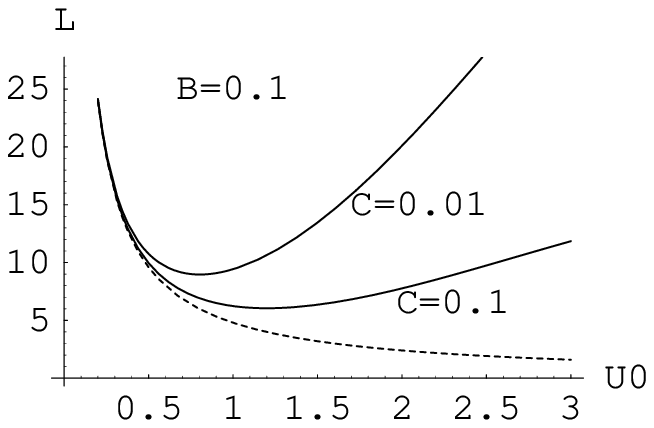}}
\\
\\
{\hspace{3cm} {\it Figure 1.  The function $L(U_0)$ for the cases of $B=0.1$ with  $C=0.01$ and $C=0.1$ respectively.  The dashed line represents that with $C=B=0$.}
\\
\\
Figure 1 shows that there presents a minimum distance between the quarks and that the minimum distance  is a decreasing function of $C$. 
\\
\\
\scalebox{1}{\hspace{5cm}\includegraphics{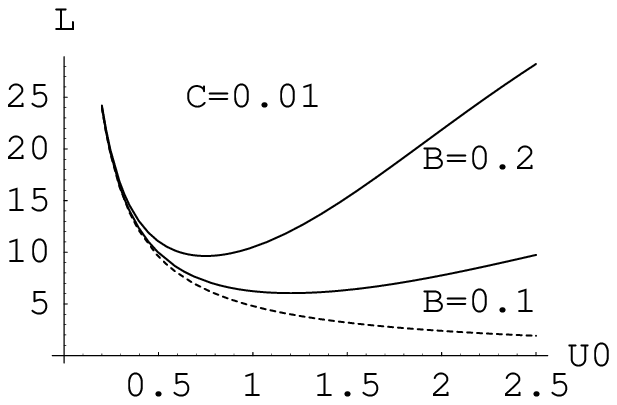}}
\\
\\
{\hspace{3cm} {\it Figure 2.  The function $L(U_0)$ for the cases of  $C=0.01$ with  $B=0.1$ and $B=0.2$ respectively.  The dashed line represents that with $C=B=0$. }
\\
\\
Figure 2 shows that there presents a minimum distance between the quarks and that the minimum distance  is an increasing function of $B$.

These properties could be read from the following analyses.    Using (3.7) we have an approximation
$$L \approx {4 (2\pi)^{3/2}\over \Gamma(1/4)^2}{1\over  U_0} + {2B^2 \Gamma(1/4)^2\over \sqrt\pi \sqrt C} U_0^2,~~~~~~if~~~~C \le B\ne0,\eqno{(3.9)}$$
which implies a minimum distance 
$$L_0 \approx {2^{2/3}4B^{2/3}\pi^{1/6}\over C^{1/2}}\left( 4\pi \over \Gamma(1/4)\right)^{2/3},~~~~~~if~~~~C \le B\ne0.\eqno{(3.10)}$$
Above result is the approximation of small values of $B$ and $C$ and shows that while the minimum value of the interquark distance $L_0$ is a decreasing function of $C$ it is an increasing function of $B$.  Note that  Eq.(3.10) implies that $L_0 \rightarrow \infty$ as $C \rightarrow 0$.  This is because that  when $C \rightarrow 0$ the string could not be localized near the boundary and we need to consider a moving string configuration as that analyzed in our previous paper [16].  Thus the above result is not an analytic function at $C=0$.

  Although the previous paper [16] had also found that the space noncommutativity will lead the interquark distance $L$ to be larger than $L_0$ the mechanism therein and that in this paper does not have a similar origin.   In [16] we consider the case of $C=0$ and find that, contrast to the figure 1, the distance $L$ is a decreasing function of $U$.  It then asymptotically approach to the minimum distance $L_0$.  The property of asymptotically approaching to the minimum distance $L_0$ means that it will exist an extremely repulsive when the quark distance is approaching to $L_0$.  {\it The repulsive force is coming from the space non-commutativity.}   Figures 1 and 2, however, show that distance $L$ could become $L_0$ at finite value of $U_0$, which means that mechanism of producing a minimum distance in this paper is not coming from the space non-commutativity.   In fact  it arises from the Melvin field which render the theory to be in a deformed background and {\it geometry of the background plays a role to produce this minimum distance $L_0$}. 

 To explicitly see  this property we may analyze the system of $B=0$.   In this case we can plot the function $L(U_0)$ by performing the numerical evaluation of (3.7).   The result is shown in figure 3 for the cases of  $C=0, 1$ and $2$ respectively. 
\\
\\
\scalebox{1}{\hspace{5cm}\includegraphics{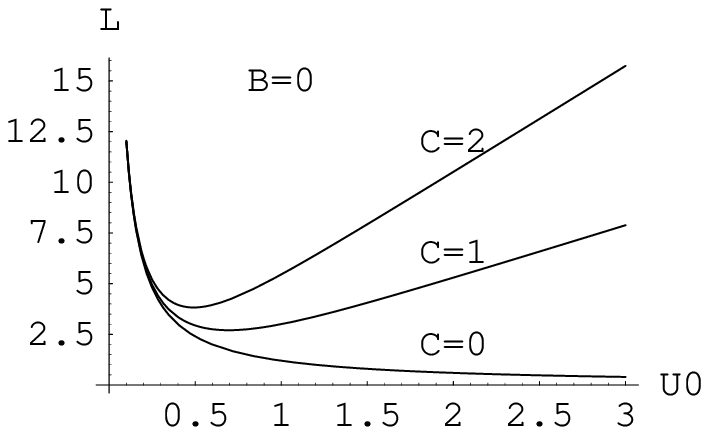}}
\\
\\
{\hspace{3cm} {\it Figure 3.  The function $L(U_0)$ for the cases of $B=0$ with  $C=0, 1$ and $2$ respectively.}
\\
\\
It then see that the distance $L$ becomes $L_0$ at finite value of $U_0$, as that in figure 1.   However, contrast to  the figure 1, the minimum distance  is an increasing function of $C$.  These properties could be read from the following analyses.  In the case of $B=0$  Eq.(3.7) gives
$$L \approx {4 (2\pi)^{3/2}\over \Gamma(1/4)^2}{1\over  U_0} + {4 \Gamma(1/4)^2\over 3\sqrt \pi}~C^{3/2}~U_0^2 ,\eqno{(3.11)}$$
which implies a minimum distance 
$$L_0 \approx {8~\Gamma(1/4)^2 \over 3 \sqrt \pi} C^{1/2}.\eqno{(3.12)}$$
The minimum distance  is an increasing function of $C$.  Thus we conclude that the mechanism behind producing a minimum distance is coming from geometry of the Melvin field deformed background and is not coming from the space non-commutativity.

   It is worthy to mention that, although (3.10) shows that the minimum distance  is a decreasing function of $C$  it is only an good approximate for $B \ge C$.  In fact, when $C$ is larger than $B$ the minimum distance will be an increasing function of $C$.  The behavior is like that in figure 3 and could be seen from the following  analyses.  In the case of $C\gg  B$ Eq,(3.7) gives an approximation  
$$L \approx {4 (2\pi)^{3/2}\over \Gamma(1/4)^2}{1\over  U_0} + {4 \Gamma(1/4)^2\over 3\sqrt \pi}~C^{3/2}~U_0^2 + {4 (2\pi)^{3/2}\over \Gamma(1/4)^2} B^2 U_0^3 , ~~~~~if~~~C \gg B, \eqno{(3.13)}$$
which implies a minimum distance 
$$L_0 \approx {8~\Gamma(1/4)^2 \over 3 \sqrt \pi} C^{1/2} + {3\over \sqrt 2} \left({4 (2\pi)^{3/2}\over \Gamma(1/4)^2}\right)^{4/3}{B^2\over C}, ~~~~~if~~~C \gg B. \eqno{(3.14)}$$
Thus the minimum distance  is an increasing function of $C$.  Note that when $B=0$ then (3.13) and (3.14) reduce to (3.11) and (3.12) respectively.

 Finally,   from figures 1, 2 and 3 we see that above $L_0$ there  have two values of $U_0$ which could correspond to a distance $L$. In fact,  the solution of large value $U_0$ will have larger energy than that of  small  value $U_0$, as analyzed in bellow.

\subsection{Analyses: Interquark Potential}
Using (3.8) we have approximations
$$H \approx - {\sqrt{2\pi}\over \Gamma(1/4)^2}\left[{4 (2\pi)^{3/2}\over \Gamma(1/4)^2} {1\over L}+ {2 \Gamma(1/4)^2\over \sqrt \pi}\left({4 (2\pi)^{3/2}\over \Gamma(1/4)^2}\right)^3{B^2\over \sqrt C}{1\over L^4}\right],  ~~~~~if~~~C \le B. \eqno{(3.15)}$$
$$H \approx - {\sqrt{2\pi}\over \Gamma(1/4)^2}\left[{4 (2\pi)^{3/2}\over \Gamma(1/4)^2} {1\over L}+ {4 \Gamma(1/4)^2\over 3 \sqrt \pi} {C^{3/2}\over L^3}+ \left({4 (2\pi)^{3/2}\over \Gamma(1/4)^2}\right)^5 {B^2\over \sqrt L^5}\right],  ~~if~~C \gg B. \eqno{(3.16)}$$
To obtain the final result we have used the relations (3.9) and (3.13) respectively.   The Melvin field will therefore slightly modify the Coulomb potential in IR.   

In the case of large $U_0$ Eqs.(3.7) and (3.8) imply the relations
$$L \approx {2\Gamma(1/4)^2\over \sqrt{2\pi}} \sqrt{C^2+B^2} U_0,\hspace{3.5cm}\eqno{(3.17)}$$
$$H \approx {\Gamma(1/4)^2\over 12 \pi\sqrt{2\pi}} C U_0^3 \approx {1\over 48 \Gamma(1/4)^4}{C\over (C^2+B^2)^{3/2}} L^3.\eqno{(3.18)}$$
Thus the   the solution of large value $U_0$ will have larger energy than that of  small  value $U_0$.  Finally,  we show in figure 4 the interquark potential $H(L)$ of the non-commutative gauge theory on the Melvin-field deformed spacetime with non-constant non-commutativity.   
\\
\\
\scalebox{1}{\hspace{5cm}\includegraphics{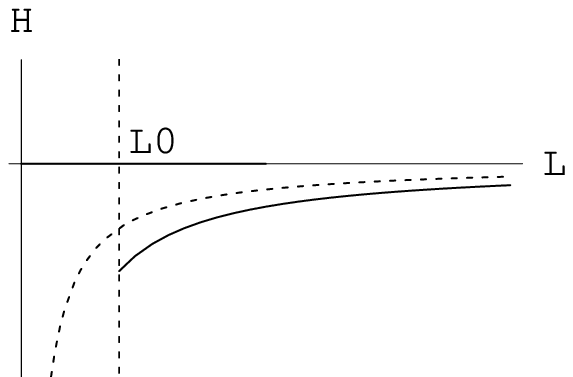}}
\\
\\
{\hspace{3cm} {\it Figure 4.  The interquark potential $H(L)$.  We see that there presents a minimum distance $L_0$ between the quarks and it becomes Coulomb phase potential in IR.  The dashed line represents that with $C=B=0$.}
\\
\\

  According to the AdS/CFT duality the property of gauge theory  is calculated from the dual string on the AdS space.   The geometric property of the AdS shall be relevant to the character the gauge theory.  As the string corresponding to the Wilson loop lies along a geodesic with endpoints on the $AdS_5$ boundary the geometry of the  $AdS_5$ therefore will bend the string to be a ``U'' type.   For the gauge theory under the Melvin field deformation the corresponding geometry is the Melvin field deformed AdS spacetime and the geometry of the deformed  $AdS_5$ will bend the string to be a ``U'' type, which in our model has shown a property of existing a minimum distance $L_0$ between the quarks.   Therefore, it is interesting to see whether such a  property  will also be shown in the trajectory of a particle.   In the section IV we will investigate this problem and see that the effect of the Melvin field on the particle is, more or less, similar to that on the string.  

\section{Particle Trajectory on Melvin Field Deformed AdS}
In this section we will analyze the motion of a particle under the  Melvin field deformed  spacetime described in (2.13).  As discussed in section III  the minimum distance $L_0$ in our model is produced by the  geometry of the background and not from the non-commutativity,  therefore we will analyze the particle trajectory on the background with $B=0$ and see what difference between that with  $C=0$ and that with  $C \ne 0$.

 We will consider the following metric
$$ds_{10}^2 =\sqrt{1+ C^2 U^4} \left[U^2\left(- dt^2+ {dx^2\over 1+ C^2 U^4} \right) + U^{-2}dU^2\right].\eqno{(4.1)}$$ 
in which $x$ is used to denote the coordinate $z$ in (2.13).
\subsection { Particle Trajectory on Undeformed AdS background}
   Let us first analyze the undeformed spacetime with $C=0$.  The equations of motion, $0=\ddot {X}^\mu + \Gamma^\mu_{\alpha \beta} \dot{X}^\alpha \dot{X}^\beta$, lead to the three equations
$$ 0 = \ddot {t} +  {2\over U}~\dot {t}~\dot {U},\hspace{3cm}\eqno{(4.2)}$$
$$ 0 = \ddot {x} +  {2\over U}~\dot {x}~\dot {U},\hspace{3cm}\eqno{(4.3)}$$
$$ 0 = \ddot {U} +   U^3\dot {t}^2 -U^3 \dot {x}^2-U^{-1}\dot {U}^2,~\eqno{(4.4)}$$
in which $\dot {t}\equiv dt/d\tau$ and $\tau$ is the proper time, and so on.  In the case of $B=0$ Eq.(4.1) implies 
$$ 1 = - U^2 \dot {t}^2 + U^2~\dot {x}^2 + U^{-2} \dot {U}^2.\eqno{(4.5)}$$
Equations (4.2) and (4.3) give the solutions
$$ \dot {t} = (\dot {t})_0 \left({U_0\over U}\right)^2,\eqno{(4.6)}$$
$$ \dot {x} = (\dot {x})_0 \left({U_0\over U}\right)^2,\eqno{(4.7)}$$
in which $U_0 \equiv U(\tau=0)$, and so on.   Substituting  (4.6) and (4.7) into  (4.5) we have a  relation
$$ \dot {U} =  \sqrt{(\dot {t})_0^2~U_0^2- (\dot {x})_0^2~U_0^2 +U^2}.\eqno{(4.8)}$$
From (4.7) and (4.8) we can find an useful differential equation
$${dx\over dU} = {(\dot {x})_0  U_0^2\over \sqrt{(\dot {t})_0^2~U_0^2  - (\dot {x})_0^2~U_0^2 +U^2 }}~{1\over U^2},\eqno{(4.9)}$$
which has a simple solution
$$ x(U)-x_0= {(\dot {x})_0  \over (\dot {t})_0^2  - (\dot {x})_0^2 }~\left({\sqrt{(\dot {t})_0^2  - (\dot {x})_0^2 + 1 }}-{\sqrt{(\dot {t})_0^2~U_0^2  - (\dot {x})_0^2~U_0^2 +U^2 }\over U}\right).\eqno{(4.10)}$$
Above solution tells us that the particle trajectory  on the AdS spacetime is  ``bended''.  For a clear illustration we show in figure 5 the particle trajectory on AdS.  
\\
\\
\scalebox{1}{\hspace{5cm}\includegraphics{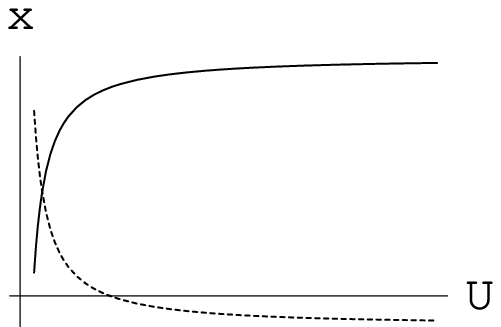}}
\\

~\hspace{1cm}{{\it Figure 5.  The particle trajectory on AdS.  Note that $x(U \rightarrow \infty) \rightarrow  a~finite~constant$. }
\\
\\
It shall be noticed that we do not want to show the particle trajectory  will be bent into a ``U" type.  What we attempt to see is the property that the ``bending'' property shown in the particle trajectory is coming from the geometric character of the background AdS spacetime which is that to bend the string to lies along a geodesic with a ``U'' type.    In fact, from the figure 3 we see that a particle trajectory along a dashing line will approach to $ x\rightarrow \pm\infty$.  Thus there is a spiky on the cross point of the dashing line and solid line. However, the string geodesic is  a smooth ``U'' type without any spiky.

\subsection { Particle Trajectory on Deformed AdS background}
We next  analyze the deformed spacetime with $C \ne 0$.  In the case of $C \gg 1$ The equations of motion become
$$ 0 \approx \ddot {t} +  {4\over U}~\dot {t}~\dot {U},\hspace{5cm}\eqno{(4.11)}$$
$$ 0 \approx \ddot {x} ,\hspace{6.7cm}\eqno{(4.12)}$$
$$ 0 \approx \ddot {U} +  2 U^3\dot {t}^2.\hspace{5cm}\eqno{(4.13)}$$
In this case (4.1) implies the relation
$$ 1 \approx - CU^4 \dot {t}^2 + {1\over C}~\dot {x}^2 + C\dot {U}^2.\eqno{(4.14)}$$
Equations (4.11) and (4.12) give the solutions
$$ \dot {t} = (\dot {t})_0 \left({U_0\over U}\right)^4.\eqno{(4.15)}$$
$$ \dot {x} = (\dot {x})_0. \hspace{1.3cm}\eqno{(4.16)}$$
Substituting  (4.15) and (4.16) into  (4.14) we have a relation
$$ \dot {U} \approx \sqrt{{1\over C}\left[1+ {C(\dot {t})_0^2 U_0^8 \over U^4} \right]}.\eqno{(4.17)}$$
From (4.16) and (4.17) we can find a useful differential equation
$${dx\over dU} = {(\dot {x})_0  \sqrt C\over \sqrt{1+ C (\dot {t})_0^2 U_0^8 U^{-4}}},\eqno{(4.18)}$$
which implies  the following relation
$$ x(U)-x_0=  \sqrt C~  \int_{U_0}^U {(\dot {x})_0~dU\over \sqrt{1+ C (\dot {t})_0^2 U_0^8 U^{-4}}}.\hspace{3cm}\eqno{(4.19)}$$
Thus
 $$x(U \rightarrow \infty) \rightarrow \infty, \eqno{(4.20)}$$
contrast to that in the undeformed system in which $x(U \rightarrow \infty) \rightarrow  a~finite~constant$.  This means that the geometry of the Melvin field deformed space has an incline to increasing the distance of final position $x(U\rightarrow \infty)$ from the initial position $x_0$.  Note that this property comes from the geometry of the Melvin field deformed background and one may conjecture that it will also be shown in the string configuration which corresponding to the Wilson loop lies along a geodesic with endpoints on the $AdS_5$ boundary.  We may argue that the corresponding property  is that shown in (3.12), i.e. interquark minimum distance $L_0$ is an increasing function of $\sqrt C$.  
\section{Conclusion}
In this paper we study the Wilson loop in the Melvin field deformed $AdS_5\times S^5$ background with NS-NS B field from the supergravity side of the duality proposed by Maldacena [4].  We first construct the background through a series of Melvin twist and a T-dual transformation on the 11D M-theory after the Kaluza-Klein reduction.  In the AdS/CFT correspondence it describes the non-commutative gauge theory  in the Melvin field deformed spacetime.  We investigate the Wilson loop therein by dual string description.  We have found that, contrast to that in the undeformed space, which was investigate by us in a previous paper [16], the string could be localized near the boundary.  Our result show that there presents a minimum distance between the quarks.  We argue that the mechanism of producing a minimum distance is not coming from the space non-commutativity but form the geometry of the Melvin field deformed background. We also have discussed the particle trajectory on the Melvin field deformed $AdS_5\times S^5$ background and see that the property of the Melvin field effect on the Wilson loop is, more or less, similar to that on the particle trajectory. 

Finally, it is known that, Maldacena method [4] cannot obtain the subleading  corrections which arise when one considers coincident Wilson loops, multiply wound Wilson loops or Wilson loops in a higher dimensional representation. In recent, Drukker and Fiol [22] showed a possible way  to compute a class of these loops using D branes carrying a large fundamental string charge dissolved on their worldvolume pinching off at the boundary of the $AdS$ on the Wilson loop [23].  It is interesting to use the D-brane approach to evaluate the  Melvin field deformed Wilson loop of other class.  Another interesting phenomena in the AdS picture of Wilson loops is the Gross-Ooguri phase transition [24] which occurs in the two Wilson loop correlator [25-27]. It would be interesting to see how the  Melvin field would affect the Gross-Ooguri phase transition therein.  These problems remain in the future investigations.
~
\\
~
\\
\begin{center} {\bf  \Large References}\end{center}
\begin{enumerate}
\item J.~M. Maldacena, ``The large N limit of superconformal field theories  and supergravity,''  Adv. Theor. Math. Phys.  2  (1998) 231-252  [hep-th/9711200].
\item E.~Witten, ``Anti-de Sitter space and holography,'' Adv.\ Theor.\ Math.\ Phys.\   2 (1998) 253 [hep-th/9802150].
\item S.~S.~Gubser, I.~R.~Klebanov and A.~M.~Polyakov, ``Gauge theory correlators from non-critical string theory,'' Phys.\ Lett.\ B 428 (1998) 105
[hep-th/9802109].
\item J.~M. Maldacena,  ``Wilson loops in large N field theories,''  Phys.   Rev. Lett.  80 (1998) 4859-4862 [hep-th/9803002]; S.-J. Rey and J.-T. Yee,  ``Macroscopic strings as heavy quarks in large  N gauge theory and anti-de Sitter  supergravity,''   Eur. Phys. J.   C22 (2001) 379--394 [hep-th/9803001]; Y. Kinar, E. Schreiber, and J. Sonnenschein, ``$Q \bar{Q}$ Potential from Strings in Curved Spacetime - Classical Results," Nucl.Phys. B566 (2000) 103-125 [9811192]; J. Gomis and F. Passerini ``Holographic Wilson Loops," JHEP 0608 (2006) 074 [hep-th/0604007].
\item E.~Witten, ``Anti-de Sitter space, thermal phase transition, and confinement in  gauge theories,'' Adv.\ Theor.\ Math.\ Phys.\   2 (1998) 505 [hep-th/9803131]; S.-J. Rey, S. Theisen and J.-T. Yee,   ``Wilson-Polyakov Loop at Finite Temperature in Large N Gauge Theory and Anti-de Sitter Supergravity,'' Nucl.Phys. B527 (1998) 171-186 [hep-th/9803135]; A. Brandhuber, N. Itzhaki, J. Sonnenschein and S. Yankielowicz,  ``Wilson Loops in the Large N Limit at Finite Temperature,'' Phys.Lett. B434 (1998) 36-40 [hep-th/9803137]; H. Boschi-Filho, N. R. F. Braga , C. N. Ferreira, ``Heavy quark potential at finite temperature from gauge/string duality," Phys. Rev D74 (2006) 086001 [hep-th/0607038].
\item N. Itzhaki, J. M. Maldacena, J. Sonnenschein, S. Yankielowicz, `` Supergravity and The Large N Limit of Theories With Sixteen Supercharges
,'' Phys.Rev. D58 (1998) 046004 [hep-th/9802042]; O. Aharony, A. Fayyazuddin, J. Maldacena, ``The Large N Limit of   N =2,1  Field Theories from Threebranes in F-theory,'' JHEP 9807 (1998) 013 [hep-th/9806159].
\item L. Girardello, M. Petrini, M. Porrati, A. Zaffaroni, ``The Supergravity Dual of N=1 Super Yang-Mills Theory,'' Nucl.Phys. B569 (2000) 451-469 [hep-th/9909047 ]; J. Polchinski and M. J. Strasslerv,'' The String Dual of a Confining Four-Dimensional Gauge Theory,''  [hep-th/0003136 ]; J. Babington, D. E. Crooks, N. Evans,'' A Stable Supergravity Dual of Non-supersymmetric Glue
,'' Phys.Rev. D67 (2003) 066007 [hep-th/0210068 ]; G.V. Efimov, A.C. Kalloniatis, S.N. Nedelko, ``Confining Properties of the Homogeneous Self-Dual Field and the Effective Potential in SU(2) Yang-Mills Theory,'' Phys.Rev. D59 (1999) 014026 [hep-th/9806165].
\item T. Mateos, J. M. Pons, P. Talavera,  ``Supergravity Dual of Noncommutative N=1 SYM,'' Nucl.Phys. B651 (2003) 291-312 [hep-th/0209150]; E. G. Gimon, L. A. P. Zayas, J. Sonnenschein, M. J. Strassler, ``A Soluble String Theory of Hadrons,'' JHEP 0305 (2003) 039 [hep-th/0212061].
\item R. Casero, C. Nunez, A. Paredes,  ``Towards the String Dual of N=1 Supersymmetric QCD-like Theories,'' Phys.Rev. D73 (2006) 086005 [hep-th/0602027].
\item  I.~R.~Klebanov and E.~Witten, ``Superconformal field theory on threebranes at a Calabi-Yau singularity,'' Nucl.\ Phys.\ B536 (1998) 199 [hep-th/9807080].
\item I.~R.~Klebanov and A.~A.~Tseytlin, ``Gravity duals of supersymmetric
SU(N) $\times$ SU(N+M) gauge theories,'' Nucl.\ Phys.\ B 578 (2000) 123 
[hep-th/0002159].
\item I.~R.~Klebanov and M.~J.~Strassler, ``Supergravity and a confining gauge theory: Duality cascades and $\chi$SB-resolution of naked singularities,'' 
JHEP 0008 (2000) 052 [hep-th/0007191].
\item J.~M.~Maldacena and C.~N\'u\~nez, ``Towards the large N limit of pure N = 1 super Yang Mills,'' Phys.\ Rev.\ Lett.\   86 (2001) 588 [hep-th/0008001].
\item A. Hashimoto and N. Itzhaki,``Non-Commutative Yang-Mills and the AdS/CFT Correspondence," Phys.Lett. B465 (1999) 142 [hep-th/9907166]. 
\item J. M. Maldacena and J. G. Russo,`` Large N Limit of Non-Commutative Gauge Theories," JHEP 9909 (1999) 025 [hep-th/9908134]; U. H. Danielsson, A. Guijosa, M. Kruczenski, and B. Sundborg,``D3-brane Holography," JHEP 0005 (2000) 028 [hep-th/0004187]; S. R. Das and B. Ghosh,``A Note on Supergravity Duals of Noncommutative Yang-Mills Theory," JHEP 0006 (2000) 043 [hep-th/0005007].
\item Wung-Hong Huang, ``Dual String Description of  Wilson Loop in Non-commutative Gauge Theory,'' [hep-th/0701069 ]. 
\item A. Hashimoto and K. Thomas, ``Dualities, Twists, and Gauge Theories with Non-Constant Non-Commutativity," JHEP 0501 (2005) 033 [hep-th/0410123]; A. Hashimoto and K. Thomas, ``Non-commutative gauge theory on D-branes in Melvin Universes," JHEP 0601 (2006) 083 [hep-th/0511197].
\item F.~Dowker, J.~P.~Gauntlett, D.~A.~Kastor and J.~Traschen, ``The decay of magnetic fields in Kaluza-Klein theory,'' Phys.\ Rev.\ D52 (1995) 6929 [hep-th/9507143]; M.~S.~Costa and M.~Gutperle, ``The Kaluza-Klein Melvin solution in M-theory,'' JHEP 0103 (2001) 027 [hep-th/0012072]; M.A. Melvin, ``Pure magnetic and electric geons,'' Phys. Lett. 8 (1964) 65; 
\item C.~G.~Callan, J.~A.~Harvey and A.~Strominger, ``Supersymmetric string solitons,'' [hep-th/9112030]; A.~Dabholkar, G.~W.~Gibbons, J.~A.~Harvey and F.~Ruiz Ruiz, ``Superstrings And Solitons,'' Nucl.\ Phys.\ B  340 (1990) 33; G.T. Horowitz and A.~Strominger, ``Black strings and P-branes,'' Nucl.\ Phys.\ B  360 (1991) 197.
\item P. Ginsparg and C. Vafa, Nucl. Phys. B289 (1987) 414; T. Buscher, Phys. Lett. B159 (1985) 127; B194 (1987) 59; B201 (1988) 466; S. F. Hassan, ``T-Duality, Space-time Spinors and R-R Fields in Curved Backgrounds,'' Nucl.Phys. B568 (2000) 145 [hep-th/9907152].
\item N. Seiberg and E. Witten, ``String Theory and Non-commutative Geometry,''  JHEP  09 (1999) 032 [hep-th/9908142].
\item N.~Drukker and B.~Fiol, ``All-genus calculation of Wilson loops using
  D-branes,''  JHEP  02 (2005) 010 [hep-th/0501109].
\item S.~A. Hartnoll and S.~Prem~Kumar,  ``Multiply wound Polyakov loops at strong coupling,''  Phys. Rev.  D74 (2006) 026001[hep-th/0603190]; S.~Yamaguchi, ``Wilson loops of anti-symmetric representation and  D5-branes,''  JHEP 05 (2006) 037 [hep-th/0603208].
\item D.~J. Gross and H.~Ooguri, ``Aspects of large N gauge theory dynamics as  seen by string theory,''  Phys. Rev. D58 (1998) 106002 [hep-th/9805129].
\item K.~Zarembo, ``Wilson loop correlator in the AdS/CFT correspondence,''
 Phys. Lett. B459 (1999) 527-534 [hep-th/9904149];  P.~Olesen and K.~Zarembo, ``Phase transition in Wilson loop correlator  from AdS/CFT correspondence,''
[hep-th/0009210].
\item H.~Kim, D.~K. Park, S.~Tamarian and H.~J.~W. Muller-Kirsten, 
  ``Gross-Ooguri phase transition at zero and finite temperature: Two circular Wilson loop case,''  JHEP 03 (2001) 003 [hep-th/0101235]; G.~Arutyunov, J.~Plefka and M.~Staudacher, ``Limiting geometries of two circular Maldacena-Wilson loop operators,''  JHEP 12 (2001) 014 [hep-th/0111290].
\item A.~Tsuji, ``Holography of Wilson loop correlator and spinning strings,'' [hep-th/0606030]; C.~Ahn, ``Two circular Wilson loops and marginal deformations,''[hep-th/0606073]; Ta-Sheng Tai and S. Yamaguchi,``Correlator of Fundamental and Anti-symmetric Wilson Loops in AdS/CFT Correspondence,'' [hep-th/0610275].
\end{enumerate}
\end{document}